\documentclass[useAMS,usenatbib]{mn2e}

\usepackage{epsfig}
\usepackage{longtable}
\usepackage{times}

\usepackage{amsmath}
\usepackage{amssymb}
\newcommand\ion[2]{#1$\;${\scshape{#2}}}%

\def \inte {$INTEGRAL$}
\def \xmm {$XMM$-$Newton$}
\def \rosat {$ROSAT$}
\def \asca {$ASCA$}
\def \sw {$Swift$}

\def \src {RX\,J0059.2--7138}

\def \hcm {\hbox {\ifmmode $ atom cm$^{-2}\else atom cm$^{-2}$\fi}}

\def \arcsec {\hbox{$^{\prime\prime}$}}

\def \apj {ApJ}
\def \aj {AJ}
\def \apjl {ApJL}
\def \apjs {ApJS}
\def \aap {A\&A}
\def \aaps {A\&AS}

\def \pasj {PASJ}

\def \mnras {MNRAS}
\def \nat {Nature}

\newcommand{\be}{\begin{equation}}

\newcommand{\ee}{\end{equation}}

\title[The 2013 outburst of the SMC pulsar \src]{Spectral properties of the soft excess pulsar \src\ during its 2013 outburst}
\author[Sidoli et al.]{L.~Sidoli,$^{1}$\thanks{E-mail: sidoli@iasf-milano.inaf.it} N.~La Palombara,$^{1}$ P.~Esposito,$^{1, 2}$ A.~Tiengo,$^{3, 1, 4}$ S.~Mereghetti,$^{1}$ \\
$^{1}$INAF, Istituto di Astrofisica Spaziale e Fisica Cosmica, Via E.\ Bassini 15,   I-20133 Milano,  Italy   \\
$^{2}$Harvard–-Smithsonian Center for Astrophysics, 60 Garden Street, Cambridge, MA 02138, USA \\
$^{3}$IUSS, Istituto Universitario di Studi Superiori, piazza della Vittoria 15,  I-27100 Pavia, Italy \\
$^{4}$INFN, Sezione di Pavia, via A. Bassi 6, I-27100 Pavia, Italy \\
}

\begin{document}

\date{Accepted 2015 March 13.  Received 2015 March 13; in original form 2015 February 18}

\pagerange{\pageref{firstpage}--\pageref{lastpage}} \pubyear{2014}

\maketitle

\label{firstpage}

\begin{abstract}
We report  on an  X--ray observation
of the  Be X--ray Binary Pulsar \src,
performed by \xmm\ in March 2014.
The 19 ks long observation was carried out about three months after the discovery 
of the  latest outburst from this Small Magellanic Cloud transient,  
when the source luminosity was    L$_X$ $\sim$10$^{38}$~erg~s$^{-1}$.
A spin period of P$_{\rm spin}$=2.762383(5)~s was derived, corresponding 
to an average spin-up of $\dot{P}_{\mathrm{spin}} = -(1.27\pm0.01)\times10^{-12}$~s $s^{-1}$ 
from the only previous  period measurement, obtained  more than 20 years earlier. 
The time-averaged continuum spectrum (0.2--12 keV) consisted of a 
hard  power-law (photon index  $\sim$0.44) with an exponential cut-off at 
a phase-dependent energy ($\sim$20-50~keV) plus  a significant soft excess below $\sim$0.5 keV.
In  addition, several features were observed in the spectrum: 
an emission line  at 6.6 keV  from highly ionized iron,    
a broad feature at 0.9-1 keV likely due to  a blend of Fe L-shell lines,
and narrow  emission and  absorption lines consistent with transitions in highly ionized oxygen, 
nitrogen and iron visible in the high resolution RGS data (0.4-2.1~keV). 
Given the different ionization stages of the narrow line components, 
indicative of photoionization from the luminous X--ray pulsar, we argue that the 
soft excess in \src\ is produced by reprocessing of the pulsar emission 
in the inner regions of the accretion disc. 
\end{abstract}


        \section{Introduction\label{intro}}

The dominant component in the spectra of X--ray binary pulsars is usually 
well described by a hard power-law model with an exponential cut-off, 
sometimes with Fe K$\alpha$ emission lines. However, at low energies several sources show a 
significant excess over the main power-law, which has been described with 
different models \citep[see][ for an overview]{LaPalombaraMereghetti06}. 
Although various physical processes have been invoked to produce it, its origin 
is still unclear. \citet{Hickox2004} showed that the origin of the 
\textit{soft excess} is related to the source total luminosity and
that a spectral component of this type has been detected in all the X--ray pulsars with a 
sufficiently high flux and small absorption. In fact, most of the soft excess sources 
are at small distances and/or away from the Galactic plane. 
This suggests that the presence of a soft spectral component could be a very common, 
if not an ubiquitous, feature intrinsic to X--ray pulsars. 
Since the study of the soft part of the spectrum in Galactic sources is affected by the interstellar absorption 
present in the Galactic plane, the best sources  to study this feature 
are the transient pulsars in the Magellanic Clouds (MCs). 
They can reach high luminosities ($L_{X} > 10^{38}$ erg s$^{-1}$), are at well known distances,  and, thanks to the low absorption in the MCs direction, 
can provide high-statistics spectra at low energies.  
In particular, the Small Magellanic Cloud (SMC) hosts more than 100 Be/X--ray Binary systems, 
about 70 of which show periodic pulsations \citep[see, e.g.][]{Haberl+12, Sturm+13, Klus+14}.

One of the most interesting pulsars in the SMC is \src,  also known as SXP~2.76,
which was discovered in 1993 by \rosat\ \citep{Hughes1994}. A Be type star was found in its error region and proposed 
as a candidate counterpart \citep{Southwell1996}.  Optical photometry of this star revealed a periodicity of 82 days, likely
related to the orbital period of the system (\citealt{Schmidtke2006}, \citealt{Bird2012}).
The ROSAT X--ray  spectrum of \src\ was unusually soft, consisting of a low temperature black-body 
component ($kT$ = 35 eV) plus  a steep power-law  (photon index $\Gamma$=2.4). 
The presence of  a soft excess was confirmed by a simultaneous $ASCA$ observation, 
which also detected X-ray pulsations  with a period of 2.76~s \citep{Kohno2000}. 
The simultaneous analysis of the \rosat\ and \asca\ data showed that the soft component 
could be described with either emission from a thin thermal plasma ($kT$=0.37~keV) 
or a broken power-law, while
a hard  power-law ($\Gamma$=0.43)   described well the spectrum above 2 keV. 
Both components had a phase--averaged luminosity of $\sim$10$^{38}$ erg s$^{-1}$ 
(0.1--10 keV): the hard component dominated above 2 keV and showed sinusoidal pulsations 
with a pulsed fraction of $\sim$37\%, while no pulsations were detected in the soft component, which 
dominated below 1 keV \citep{Kohno2000}.

After the 1993 observations, no other bright outbursts were seen from \src\  
until 2013, except for two detections of the source with PCA $RXTE$  
in September 2002 and in November 2004 reported by \citet{Galache2008}.  
The 2013  outburst was discovered with the $Swift/BAT$ instrument, 
which detected the source starting from December 26 (see Fig.~\ref{fig:bat}; \citealt{Krimm2014}).
A couple of weeks later,  \inte\  monitoring observations  of the SMC confirmed the bright status of the source \citep{Coe2014}.  
After the discovery of this outburst, we triggered our \xmm\ Target of Opportunity program 
aimed at investigating the soft excess in  high luminosity Be/X--ray transients (hereafter Be/XRTs) 
thus obtaining an  observation of \src\  in March 2014. Here we report the results of this observation.

 	 \section{Observations and Data Reduction}
         \label{data_redu}

\begin{figure}
\begin{center}
\centerline{\includegraphics[width=6.0cm,angle=-90]{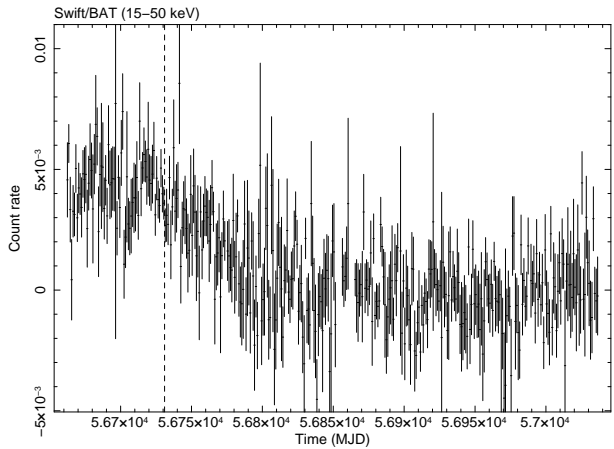}} 
\caption{\src\ outburst as observed by \sw/BAT (15--50 keV). The bin time is one day. The BAT light curve is publicly available at
{\sc http://swift.gsfc.nasa.gov/results/transients/} \citep{Krimm2013}.
The date of our  \xmm\ observation is indicated by ß the vertical dashed line.
}
\label{fig:bat}
\end{center}
\end{figure}

\src\ was observed by \xmm\ between 2014 March 14 and 15,
for a net exposure of $\sim$19~ks (see Table~\ref{tab:log} for the observation log).
In Fig.~\ref{fig:bat} we show the time of the \xmm\ observation in the context of
the outburst as observed by Swift/BAT (15--50 keV).

\begin{table*}
\caption{\label{tab:log}
         Summary of the \xmm\ observation (ID. 0674730201) of \src
}
\begin{center}
\begin{tabular}{lll@{\ }cccr@{$\pm$}l}
\hline
Instrument    & Filter
              & Mode
              & \multicolumn{2}{c}{Start and End Times}
              & Net Exp. time (ks)
              & \multicolumn{2}{c}{Net count rate (${\rm s}^{-1}$)} \\
\hline
pn            &  Thin~1  & Small Window  & 2014-03-14 19:45:57 & 2014-03-15 01:10:34 & 13.7 & 19.24  & 0.04 \\
MOS1          &  Thin~1  & Small Window & 2014-03-14 19:40:09 & 2014-03-15 01:07:44 &  19.6 &  5.92   &  0.02  \\
MOS2          &  Thin~1  & Timing       & 2014-03-14 19:40:48 & 2014-03-15 01:03:30 & 19.4  &  5.86   &  0.02 \\
RGS1          &  $-$    & Specroscopy       & 2014-03-14 19:39:58 & 2014-03-15 01:11:55 &  19.9  & 0.246 &  0.004 \\
RGS2          &  $-$    & Specroscopy       & 2014-03-14 19:40:06 & 2014-03-15 01:11:53 &  19.8  & 0.299  &  0.004 \\
\hline
\end{tabular}
\end{center}
\end{table*}

The \xmm\ $Observatory$ 
carries three 1500~cm$^2$ X--ray
telescopes, each with an European Photon Imaging Camera (EPIC; 0.2-12 keV)
at the focus. Two of the EPIC detectors use MOS CCDs \citep{Turner2001} and one uses pn CCDs
\citep{Struder2001}. Reflection Grating Spectrometer (RGS; 0.4-2.1 keV)
arrays \citep{DenHerder2001} are located 
in the telescopes with MOS cameras at their primary focus.
The EPIC cameras were operated in different modes, but always adopting the thin~1 filter:
pn in Small Window, 
MOS1 in Imaging Prime Partial Window2 (small window),
MOS2 in Timing Fast Uncompressed in the central CCD, while in Imaging Mode in the other CCDs. 
The time resolution is 5.7~ms, 0.3~s and 1.75~ms in the pn, MOS1 and MOS2, respectively.
The data were reprocessed using version 13.5 of the Science Analysis Software ({\sc sas}) with standard procedures.

The RGSs were operated in spectroscopy mode and
their data were analysed in the energy range 0.4--2.1 keV.  
The {\sc sas} task {\sc rgsproc} was used to
extract the RGS1 and RGS2 spectra, which were later combined into one single grating spectrum
using {\sc rgscombine}.

EPIC source counts were extracted from circular regions
of 40\arcsec\ radius for both the pn and MOS1, adopting PATTERN from 0 to 4 (mono- and bi- pixel events) 
in the pn 
and from 0 to 12 (up to 4-pixel events) in the MOS1.
Background counts were obtained from similar sized regions offset from the source positions. 
Due to calibration issues in the MOS2 data (see Appendix~\ref{app:m2} for details),
we decided to use only the  pn and MOS1 data for the spectral analysis,
while we used all three EPIC data sets for the timing analysis.
No further filtering based on the background level was necessary.
Appropriate response matrices were generated
using the {\sc sas}  tasks {\sc arfgen} and {\sc rmfgen}.

Free relative normalizations between the
two cameras were included to account for uncertainties in
instrumental responses.
The normalization factor for the MOS1 spectra relative to PN
(factor fixed at 1) was always around 0.965$\pm{0.007}$.
All spectral uncertainties and upper-limits are given at 90\% confidence level for
one interesting parameter.

In the spectral fitting we adopted the interstellar abundances of \citet{Wilms2000}  
and  photoelectric absorption cross-sections of \citet{bcmc},
using the absorption model  {\sc phabs} in {\sc xspec}.
We checked  that using a different absorption model ({\sc TBabs} in {\sc xspec})
with Wilms et al. (2000) abundances and cross sections by \citet{Verner1996} similar spectral fit results were obtained.

\begin{figure}
\begin{center}
\centerline{\includegraphics[width=10.50cm,angle=-90]{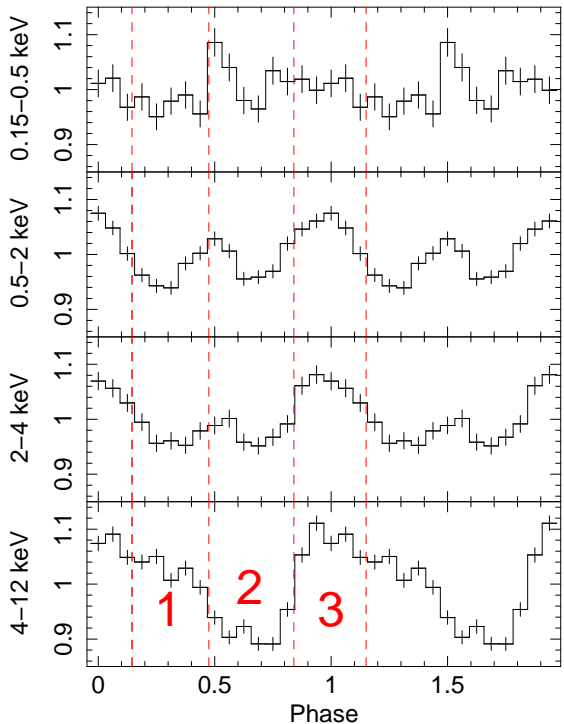}} 
\caption{\src\ pulse profiles observed with the EPIC pn camera in four energy bands. 
The vertical dashed lines and  the numbers mark the three
spin-phase intervals adopted for the  phase-selected spectroscopy (Sect.~\ref{sect:sel}).
}
\label{fig:efold_energy}
\end{center}
\end{figure}

  	\section{Analysis and Results\label{result}}

\subsection{Timing Analysis}
\label{sec:timing}

The data from the three EPIC cameras in the 0.15--12 keV range were used for the
timing analysis after conversion of their times of arrival to the Solar System barycenter.
Using the $Z^2_2$ test,  the spin period of \src\ 
was measured to be $P_{\mathrm{spin}}=2.762383(5)$~s.

Following \citet{Kohno2000}, we defined the pulsed fraction as
$(I_{\mathrm{max}}-I_{\mathrm{min}})/2I_{\mathrm{average}}$, where
$I_{\mathrm{max}}$, $I_{\mathrm{min}}$, and $I_{\mathrm{average}}$ are
the maximum, minimum, and average count rate, respectivily. The
average pulsed fraction in the 2--10~keV energy range was
$8.9\pm0.5$~\% and during the observation the value did not change more than $\sim$20\%.

With respect to the last observation performed in 1993 (MJD 49119-49120) with $ASCA$, 
when \citet{Kohno2000} found $P_{\mathrm{spin}}=2.763221(4)$~s,
our value implies 
an average spin-up rate of $\dot{P}_{\mathrm{spin}} = -(1.27\pm0.01)\times10^{-12}$~s $s^{-1}$.

\begin{figure*}
\begin{center}
\centerline{\includegraphics[width=14.5cm,angle=0]{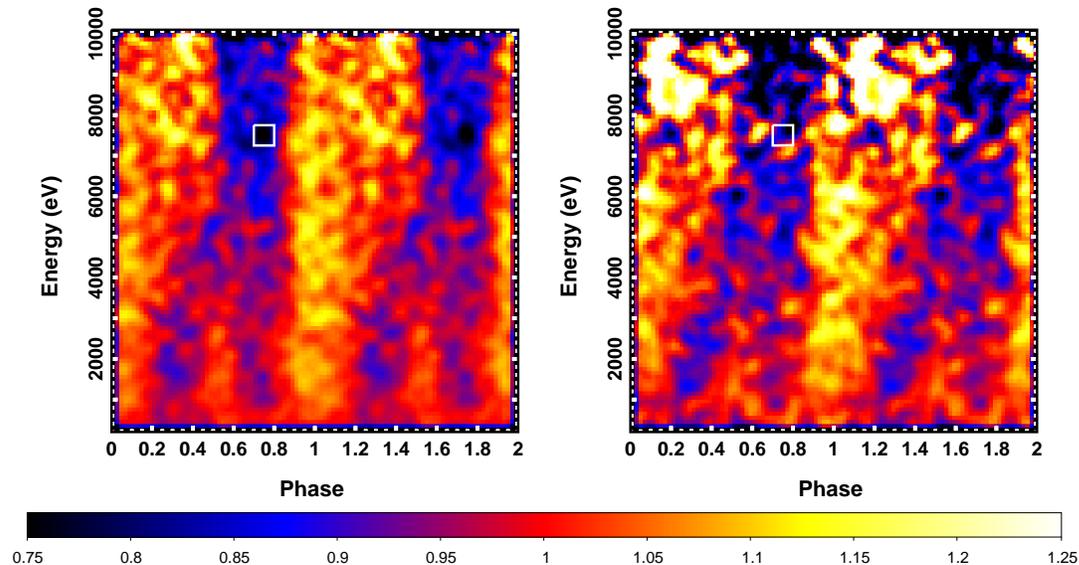}} 
\caption{\src\ phase-energy image  (normalized to the phase-averaged spectrum)
obtained from  the  EPIC pn and MOS2 data 
(the MOS1 time resolution is too poor for
such a fine phase binning). 
The small white box marks a {\em hole} in the
image, indicating a deficit of counts, with respect to the average spectrum, 
in a narrow range of energies  ($\sim$7-8 keV) and spin phases ($\Delta\phi$=0.7-0.8).}
\label{fig:efold_ima}
\end{center}
\end{figure*}

In Fig.~\ref{fig:efold_energy} we show the EPIC pn pulse profiles in different energy ranges.
The pulse shape is clearly energy-dependent, with a broader single asymmetric peak in the  4-12 keV range,
whereas a clear secondary pulse emerges at intermediate energies (0.5-4 keV). 
Below 0.5 keV the pulsations are less evident, but still significant (at 3.6~$\sigma$ confidence level). 
The pulse profile evolution with energy is clearly shown also by the phase-energy image (Fig.~\ref{fig:efold_ima}). 
This image represents the number of source counts in small phase and energy intervals, 
normalized with respect to the phase-averaged spectrum (see \citealt{Tiengo2013}).
A deficit of counts (marked by the white box) over a narrow range of energies ($\sim$7-8 keV) 
in the phase interval $\Delta\phi$=0.7-0.8 is evident. 
This possible evidence for a phase-dependent absorption line will be discussed in Sect.~\ref{sec:spec}.

\subsection{EPIC Spectroscopy}
\label{sec:spec}

\begin{table}
\begin{center}
\caption[]{Results of the  simultaneous EPIC pn and MOS1 fit of the  time-averaged spectrum (0.2--12 keV). 
The double-component continuum consists of a cut-off power law  
({\sc cutoffpl} in {\sc xspec}) and a black-body model. Both components are 
equally absorbed with column density N$_{\rm H}$. 
Three Gaussian lines in emission are needed to account for  positive residuals in the spectrum 
(the line at  E$_{line1}$ has been added only in the pn spectrum, being of instrumental origin).
The parameters of the black-body component are its temperature, kT$_{\rm bb}$,
and it radius,  ${\rm r_{bb}}$.
A distance of 61~kpc is assumed \citep{Hilditch2005}.
The cut-off power-law continuum is defined as A(E)=k${\rm E^{-\Gamma}\exp-(E/E_{c})}$, 
where $\Gamma$ is the photon index and 
E$_{c}$ is the e-folding energy of the exponential rolloff (in units of keV).
Flux$_{line\#N}$ is the total flux in the Gaussian line \#N.
EW$_{line\#N}$ is the equivalent width of the line \#N (eV).
The unabsorbed  flux is in units of erg~cm$^{-2}$~s$^{-1}$, the luminosity in units of erg~s$^{-1}$.
L$_{\rm bb}$/L$_{\rm cpl}$ is the fraction of X--ray emission (corrected for the absorption) 
contributed by the soft component
with respect to the cut-off power-law, in the broad energy band 0.01-10 keV.
}
\begin{tabular}{lc}
 \hline
\hline
\noalign {\smallskip}
Parameter                                  &     Value                            \\
\hline
\noalign {\smallskip}
N$_{\rm H}$   ($10^{20}$~cm$^{-2}$)        &  $ 2.3 ^{+0.6} _{-0.5} $             \\
\hline
kT$_{\rm bb}$ (keV)                        &  $0.093 ^{+0.005} _{-0.005}$         \\
${\rm r_{bb}}$ (km)                        &  $350^{+80} _{-50}$                  \\
\hline
$\Gamma$                                   &  $0.44 ^{+0.02} _{-0.02}$            \\
E$_{c}$ (keV)                              &  $36 ^{+8} _{-6}$                    \\
\hline
E$_{line1}$   (keV)                        &  $2.23 \pm{0.01}$                    \\ 
\hline
E$_{line2}$   (keV)                        &  $6.58 ^{+0.07} _{-0.06}$            \\
$\sigma$$_{line2}$ (keV)                   &  $0.13 ^{+0.09} _{-0.08}$            \\
Flux$_{line2}$ (ph~cm$^{-2}$~s$^{-1}$)     &  (6$\pm{2}$)$\times10^{-5}$          \\
EW$_{line2}$   (eV)                        &  $30 \pm{10}  $                      \\
\hline
E$_{line3}$   (keV)                        &  $0.96 ^{+0.02} _{-0.03}$            \\
$\sigma$$_{line3}$ (keV)                   &  $0.095 ^{+0.030} _{-0.024}$         \\
Flux$_{line3}$   (ph~cm$^{-2}$~s$^{-1}$)   &  $15 ^{+5} _{-4}$$\times10^{-5}$     \\
EW$_{line3}$   (eV)                        &  $30\pm{8} $                 \\
\hline                                                    
L$_{\rm bb}$/L$_{\rm cpl}$  (0.01-10 keV)  &   1.7\%                              \\                     
\hline
Unabsorbed flux (0.5--10 keV)              &    1.58 $\times$10$^{-10}$           \\
Luminosity  (0.5--10 keV)                  &    7$\times$10$^{37}$                \\
$\chi^{2}_{\nu}$/dof                       &    1.405/546                         \\
\noalign {\smallskip}
\hline
\label{tab:av_spec}
\end{tabular}
\end{center}
\end{table}

\subsubsection{Time-averaged spectrum}
\label{sect:av}

We searched for possible spectral variability within the \xmm\ exposure,
by extracting light curves in different energy ranges.
We found no strong evidence for intensity and spectral variability
on timescales longer than the spin periodicity, so we extracted a time-averaged spectrum for the whole  pn and MOS1 exposures.

\begin{figure}
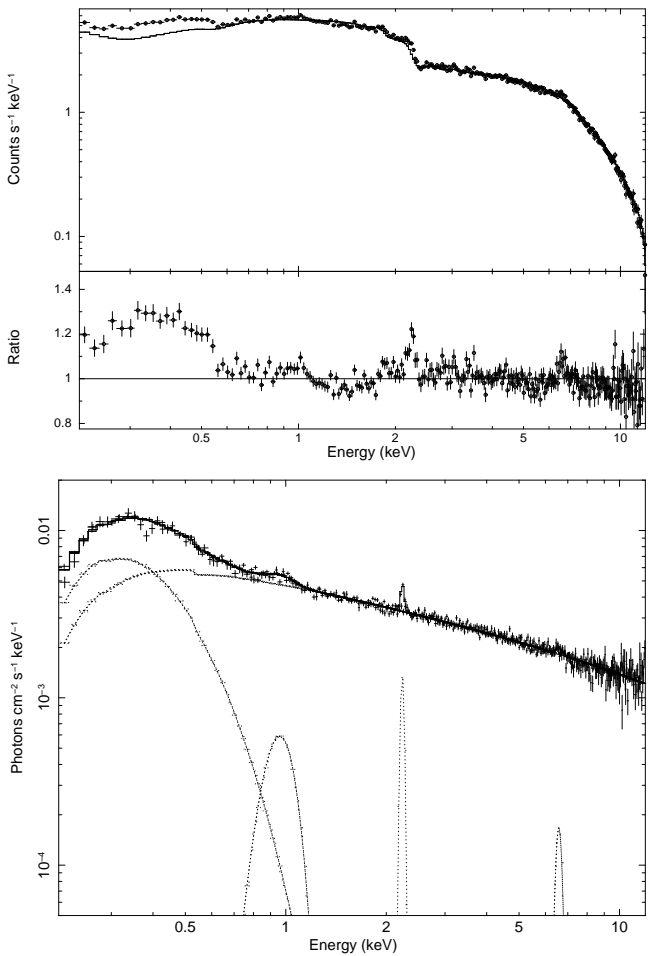

\begin{center}
\begin{tabular}{cc} 
\includegraphics[height=8.5cm,angle=-90]{./mn_fig4a.ps} \\
\includegraphics[height=8.5cm,angle=-90]{./mn_fig4b.ps}
\end{tabular}
\end{center}
\caption{\src\ EPIC time-averaged spectrum. 
{\em Upper panel} shows the fit of the pn data with a single power-law, to clearly show the main 
structures in the spectrum.
{\em Lower panel} displays the photon spectrum from the EPIC pn and MOS1 observations fitted with
the model reported in Table~\ref{tab:av_spec}. 
}
\label{fig:av_spec}
\end{figure}

The main spectral structures in addition to a simple absorbed power-law component,
 can be seen in Fig.~\ref{fig:av_spec} ({\em upper panel}):
1)-a soft excess below $\sim$0.5~keV;
2)-a high energy cut-off;
3)-a narrow feature at  2.2~keV (present only in the pn data, 
very likely due to residual calibration uncertainties around the Au edge);
as done by other authors (e.g. \citealt{DiazTrigo2014}) we model it with a Gaussian line,
but  we will not discuss it  further;
4)-an emission line at 6.5-6.8~keV;
5)-a broad positive residual around 0.9-1.0 keV.
The presence of  the latter structure also in the RGS spectrum (see below, Sect.~\ref{sec:rgs}),
excludes a calibration/instrumental origin, thus we decided to include an additional broad Gaussian line to fit it.

To account for these structures, we adopted 
a double-component continuum consisting of a cut-off power-law plus a black-body model
together with the three Gaussian lines. 
The results are reported in Table~\ref{tab:av_spec} and displayed in Fig.~\ref{fig:av_spec} ({\em lower panel}).

We also tried a hot thermal plasma model ({\sc mekal} in {\sc xspec}) instead of the black-body 
to fit the soft excess, with similar $\chi^2$~results: 
all the parameters were consistent with those reported in Table~\ref{tab:av_spec},
except for the column density N$_{\rm H}$=$(4.1\pm{0.6})\times10^{20}$~cm$^{-2}$ and
the fraction of X--ray emission contributed by the soft component with respect to the cut-off power-law,
which was 7\% (instead of 1.7\%), in the energy band 0.01-10 keV.
The parameters of the {\sc mekal} model were the following: 
a plasma temperature $kT_{\rm M}$=0.21$\pm{0.03}$~keV, 
a very low metal abundance $Z \lesssim 0.007$ $Z_{\sun}$ (too low for the SMC, see below), 
a normalization, $norm_{\rm M}$, of 0.025$^{+0.008} _{-0.006}$~cm$^{-5}$
(in our case, $norm_{\rm M}$=2.4$\times10^{-62}\int n_en_H\mathrm{d}V$, 
where $n_e$ and $n_H$ are the electron and hydrogen  densities, $V$ is the emitting volume).

We also tried to adopt two absorption models,
one to account for the total Galactic absorption in the SMC direction (N$_{\rm H}$  fixed at 6$\times$$10^{20}$~cm$^{-2}$),
the other one to account for the local absorption in the SMC ({\sc vphabs} model in {\sc xspec} 
with an abundance fixed at 0.2 for metals, appropriate for the SMC),
but we did not obtain better fits than using a single absorption model with a free column density.
So, we decided to adopt a single model for the absorption ({\sc phabs} model in {\sc xspec}).
We  also tried to fit the data with  warm and ionized absorbers 
(like {\sc absori} in {\sc xspec}; \citealt{Done1992}) 
in addition to a neutral one accounting for the interstellar absorption, but this did not result in better fits.

\subsubsection{Spin-phase-selected spectroscopy}
\label{sect:sel}

Given the  energy-dependence observed in the folded 
spin profiles (Figs.~\ref{fig:efold_energy} and \ref{fig:efold_ima}), 
it is not surprising that the fits to the time-averaged spectrum were not formally acceptable.
To quantitatively explore the observed spectral evolution,
we extracted three  phase-selected spectra 
(marked by numbers in Fig.~\ref{fig:efold_energy})
for EPIC MOS1 and the pn: 
$\Delta\phi$=0.15-0.45 (spectrum n.~1),
$\Delta\phi$=0.45-0.85 (spectrum n.~2),
$\Delta\phi$=0.85-0.15 (spectrum n.~3).

The MOS1 and pn spectra were fitted simultaneously in the 0.2 to 12\,keV range with 
the same model used for  the time-averaged  emission.
Good fits were obtained  with the spectral parameters listed in Table~\ref{tab:bbpow}.
Similar values were obtained using a {\sc mekal} model instead of the black-body,
but again with a very low metallicity  ($Z \lesssim 0.01$ $Z_{\sun}$).

The results reported in Table~\ref{tab:bbpow} indicate  that the phase-dependent spectral variability  
is mainly due to variations in the high energy cut-off.
The iron lines around $\sim$6.6~keV are significant 
at 2.9~$\sigma$, 4.4~$\sigma$ and 4.1~$\sigma$, in spectra~1, ~2 and ~3, respectively.
The significance was estimated by increasing the $\Delta\chi^2$ value until 
the confidence region boundary of the Gaussian normalisation crosses 0.
The iron line centroid energy shows variations from 6.77~keV (in spec.~1) to 6.54~keV (in spec.~2).
In spec.~2 the centroid energy is compatible with transitions in \ion{Fe}{xxi}-\ion{Fe}{xxii} ions
\citep{Kallman1982}, in spec.~3 with  \ion{Fe}{xxi}-\ion{Fe}{xxv} ions, while in spec.~1 (line centroid at 6.77~keV),
it is not consistent (within the 90\% errors) with any known line.
Although the emission line in spec.~1 is significant at less than 3~$\sigma$, if we identify it with 
 emission  from  \ion{Fe}{xxv} at 6.7~keV, a hint for a blue-shift of 900~km~s$^{-1}$ is present,
possibly indicating an outflowing photoionized plasma.
Also the flux of the broad feature at 0.9-1 keV is variable along the spin profile, while
the other spectral parameters are constant, within the uncertainties.

\begin{table*}
\begin{center}
\caption[]{Results of the  spin-phase selected spectroscopy (spectra n. 1, 2 and 3 
indicate spin phase intervals 0.15-0.45, 0.45-0.85, 0.85-0.15 respectively) 
fitting together the EPIC pn and MOS1 observations in the energy band (0.2--12 keV). 
The model is the same adopted in Table~\ref{tab:av_spec}.
}
\begin{tabular}{lccc}
 \hline
\hline
\noalign {\smallskip}
Parameter                &            1                            &     2                                      &        3                                 \\
\hline
\noalign {\smallskip}
N$_{\rm H}$ ($10^{20}$~cm$^{-2}$) &  $ 2.9 ^{+1.9} _{-1.3}$        &    $2.1  ^{+0.9} _{-0.8}$                  &    $1.9  ^{+1.2} _{-1.0}$                 \\
\hline
kT$_{\rm bb}$ (keV)     &  $0.09 ^{+0.01} _{-0.02}$                &   $0.099  ^{+0.008} _{-0.007}$             &     $0.09  ^{+0.01} _{-0.01}$            \\
${\rm r_{bb}}$ (km)     & $420^{+700} _{-130}$                     &     $300^{+100} _{-60}$                    &     $350^{+230} _{-90}$                \\
\hline
$\Gamma$                &  $0.43 ^{+0.04} _{-0.05}$                &   $0.42 ^{+0.03} _{-0.03}$                 &     $0.37 ^{+0.04} _{-0.04}$              \\
E$_{c}$ (keV)           &  $>$45                                   &   $20  ^{+4} _{-3}$                        &     $22  ^{+5} _{-4}$                      \\
\hline
E$_{line1}$   (keV)     &  $2.23 ^{+0.03} _{-0.03}$                &   $2.23 ^{+0.02} _{-0.02}$                 &     $2.25 ^{+0.03} _{-0.03}$              \\ 
\hline
E$_{line2}$   (keV)     &  $6.77 ^{+0.06} _{-0.04}$                &   $6.54 ^{+0.05} _{-0.04}$                 &     $6.59 ^{+0.08} _{-0.08}$              \\
$\sigma$$_{line2}$ (keV)&   $<$1.6                                 &   $<$0.09                                  &     $0.12 ^{+0.09} _{-0.08}$              \\
Flux$_{line2}$   (ph~cm$^{-2}$~s$^{-1}$)         &  (4.3$\pm{2.4}$)$\times10^{-5}$          &   (4.9$ ^{+2.7} _{-1.8}$)$\times10^{-5}$   &     (8.4$ ^{+3.9} _{-3.6}$)$\times10^{-5}$     \\
EW$_{line2}$   (eV)     &  $23\pm{13}$                             &        $30  ^{+16} _{-10}$           &       $43 ^{+20} _{-18}$                 \\
\hline
E$_{line3}$   (keV)     &  $0.92 ^{+0.07} _{-0.17}$                &   $0.99 ^{+0.03} _{-0.04}$                 &     $0.93 ^{+0.05} _{-0.05}$              \\
$\sigma$$_{line3}$ (keV)&    $0.19 ^{+0.09} _{-0.06}$              &   $<$0.09                                  &        $0.10 ^{+0.06} _{-0.04}$            \\
Flux$_{line3}$     (ph~cm$^{-2}$~s$^{-1}$)       &   $(40 ^{+50} _{-20})$$\times10^{-5}$    &    $(9 ^{+5} _{-4})$$\times10^{-5}$        &       $(16 ^{+14} _{-7})$$\times10^{-5}$        \\
EW$_{line3}$   (eV)     &  $80   ^{+80} _{-30} $                   &   $18  ^{+11} _{-8} $                      &       $31 ^{+23} _{-14}$                    \\
\hline
L$_{\rm bb}$/L$_{\rm cpl}$  (0.01-10 keV)    &             0.02       &          0.017                          &             0.015                  \\         
\hline                                                    
Unabsorbed flux   (0.5-10 keV) &    1.6    $\times10^{-10}$       &     1.4$\times10^{-10}$                    &    1.7$\times10^{-10}$              \\
\hline
$\chi^{2}_{\nu}$/dof           &    1.041/517                     &     1.067/522                              &    1.033/518                         \\
\noalign {\smallskip}
\hline
\label{tab:bbpow}
\end{tabular}
\end{center}
\end{table*}

The phase-energy image (Fig.~\ref{fig:efold_ima}) shows  a hint for a deficit of counts at $\sim$7--8~keV, with respect to the spin-averaged spectrum.
This deficit, possibly indicating an absorption feature, occurs in a narrow range of spin phases ($\phi$$\sim$0.7--0.8).
We extracted a spectrum corresponding to this phase interval and we fitted it adding a Gaussian line in absorption.
The new fit  gave a reduced $\chi^{2}_{\nu}$=1.014 for 464 dof 
(compared to $\chi^{2}_{\nu}$=1.049 for 467 dof with no line).
The resulting line parameters were the following:
centroid energy E$_{line4}$=$7.50 \pm{0.09}$~keV,
width $\sigma$$_{line4}$$<0.27$~keV,
total Flux$_{line4}$=$-(1.28^{+0.62}_{-0.79})$$\times10^{-4}$~photons~cm$^{-2}$~s$^{-1}$,
equivalent width EW$_{line4}$$\sim$$-$80~eV. 
The single trial significance of this line is 4.4~$\sigma$, but given that many phase intervals
were explored, the overall significance is too small to claim a firm detection.

\subsection{RGS Spectroscopy}
\label{sec:rgs}

\begin{figure}
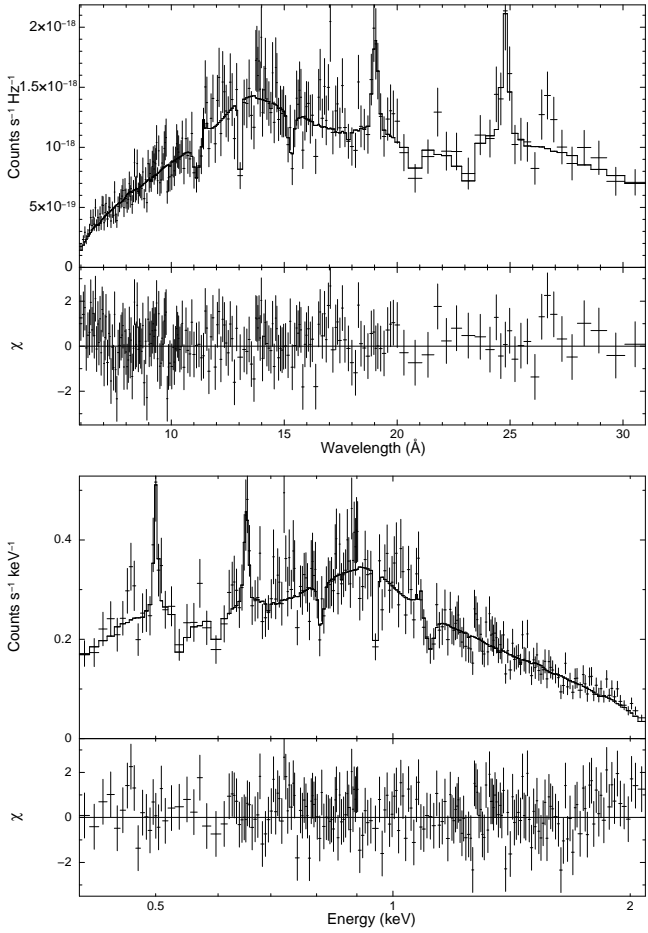

\begin{center}
\begin{tabular}{cc}
\includegraphics[height=8.5cm,angle=-90]{mn_fig5a.ps} \\
\includegraphics[height=8.5cm,angle=-90]{mn_fig5b.ps}
\end{tabular}
\end{center}
\caption{Results of the RGS spectroscopy. The counts spectrum is shown
when fitting the combined RGS1 and RGS2 spectra
with a double component continuum (a black-body together with a power-law) modified by the absorption,
including six Gaussian lines to account for the brightest lines 
(four in emission and two in absorption; see 
Table~\ref{tab:rgslines}).  In the {\em upper panel} the spectrum is in units of wavelength,
while in the {\em lower panel} it is in units of energy. 
Spectra have been severely rebinned in these plots, to better show the residuals.
}
\label{fig:rgs}
\end{figure}

The RGS first order spectra (0.4--2.1 keV) were extracted from the whole exposure
and combined into a single grating spectrum. 
We fitted it with the same double-component model resulting from the time-averaged spectrum (black-body plus cut-off power-law).
The inspection of the residuals clearly showed the presence of narrow lines (both in absorption and in emission), 
together with a broad structure around 0.9-1.0 keV (which was also clearly present in the EPIC spectra).
The addition of Gaussian lines to account for them, resulted in the
line parameters reported in Table~\ref{tab:rgslines}.  The spectrum is shown in Fig.~\ref{fig:rgs}. 
The two brightest narrow lines, at 0.5 and 0.65 keV, are firmly  associated with \ion{N}{vii} and \ion{O}{viii} Ly$\alpha$ lines,
respectively, while the broad component at 0.92~keV is possibly due to different contributions: 
a blend
of several emission lines from iron in a range of ionizations states, from   \ion{Fe}{xviii} to \ion{Fe}{xx}, 
a radiative recombination continuum (RRC) from \ion{O}{viii}, likely with superposed weak emission lines 
from ionized neon (Table~\ref{tab:rgslines}).
The parameters of this broad component observed in the RGS spectrum are consistent with what found with EPIC.

We then used a thermal plasma model ({\sc mekal}) with free abundance instead of the black-body
to account for the soft excess (with the metal abundance let free to vary,  and resulting in 
a kT of $\sim$0.23 keV), but in this case only the narrow line at 0.65 keV could be well fitted.
Also adopting additional {\sc mekal} models with different temperatures or a {\sc cemekl} model 
(a multi-temperature plasma emission model built from the mekal code, 
with emission measures following a power-law in temperature)
we could not obtain acceptable fits to the other emission lines, likely indicative  of a photoionized nature of the emitting matter.
Two absorption lines were also found, with possible identifications listed in Table~\ref{tab:rgslines}.

The \ion{O}{vii} He$\alpha$ triplet is undetected, preventing us from using the 
so-called ``G ratio'' in He-like ions
to obtain information on the ionization process and plasma density \citep{Porquet2000}.
We could place the following upper limits (90\% confidence level) to the fluxes of
the three  \ion{O}{vii} triplet components:  
F$<$3.1$\times10^{-5}$~${\rm ph}~{\rm cm}^{-2}{\rm s}^{-1}$ (forbidden line at 0.561~keV),
F$<$5.0$\times10^{-5}$~${\rm ph}~{\rm cm}^{-2}{\rm s}^{-1}$ (intercombination line at 0.569~keV),
F$<$3.4$\times10^{-5}$~${\rm ph}~{\rm cm}^{-2}{\rm s}^{-1}$ (resonance line at 0.574~keV).

\begin{table*}
\begin{center}
\caption[]{Main lines identified in the \src\ RGS spectrum (0.4-2.1 keV). Negative fluxes and negative equivalent widths (EWs) mean that the Gaussian lines are in absorption.}
\begin{tabular}{cclccrr}
 \hline
\hline
\noalign {\smallskip}
Energy$_{\rm observed}$        &            Ion     &    $\lambda_{\rm lab}$    &    Energy$_{\rm lab}$       &   $\sigma$              &   Flux                                              &  EW \\
(keV)                         &                    &           ($\text{\AA}$)  &     (keV)  & (eV)                 &  ($10^{-5}~{\rm ph}~{\rm cm}^{-2}{\rm s}^{-1}$) &  (eV)  \\
\hline
\noalign {\smallskip}
0.500  $^{+0.001} _{-0.002}$  &   \ion{N}{vii}     &      24.779    (Ly$\alpha$) &     0.50                  &   2.4   $^{+2.0} _{-1.0}$ &   6.8$\pm{2.7}$                  &   7 $\pm{3}$   \\
0.6516  $^{+0.0026} _{-0.0016}$ &   \ion{O}{viii}  &      18.967    (Ly$\alpha$) &     0.65                  & 4   $^{+2} _{-1}$         &   5.7$\pm{2.0}$                  &   9 $\pm{3}$   \\
0.811  $^{+0.003} _{-0.007}$  &   \ion{Fe}{xvii}  (?)   & 15.2610                &     0.81                  & 5  $^{+7} _{-2}$         &   $-$2.8$^{+1.2} _{-1.5}$         &   $-$5$\pm{2}$   \\
0.921  $^{+0.026} _{-0.024}$  &   \ion{Fe}{xviii}-\ion{Fe}{xx} (blended?) &  $-$ &     $-$                   &  60  $^{+30} _{-40}$     &   17$^{+5} _{-6}$                 &   34$^{+20} _{-14}$    \\
                              &   or RRC from \ion{O}{viii} (?)-- \ion{Ne}{ix} (?)  & 13.45 &  0.89; 0.91    &                          &                                   &   \\
1.0795  $^{+0.0009} _{-0.0056}$  &  \ion{Fe}{xxii} (?) &    11.490               &       1.079               & $<$9                     &   4.7$^{+2.6} _{-2.4}$            &   11 $\pm{6}$   \\
1.114  $^{+0.013} _{-0.012}$  &  \ion{Fe}{xxiii} (?) &    11.019              &        1.125              & 9   $^{+21} _{-6}$       &   $-$4.4$^{+1.9} _{-2.8}$         &   $-$9$^{+4} _{-5}$  \\
\noalign {\smallskip}
\hline
\label{tab:rgslines}
\end{tabular}
\end{center}
\end{table*}

	      \section{Discussion}\label{sec:discussion}

The Be transient \src\ in the SMC  attracted attention  because, like a few other  X--ray pulsars, it showed a prominent soft excess, 
the origin of which is still uncertain \citep{Hickox2004}.
Two were the main alternatives discussed by Hickox et al. to explain the soft excess in \src:
either reprocessing of X--rays in the inner region of an accretion disc or  optically thin, hot thermal emission.

Thanks to  the large collecting area of \xmm,  the observation obtained during the recent new outburst of \src, 
allowed us to perform a more detailed study of this source,  
even if its luminosity  of L$_{\rm X}$=7$\times$10$^{37}$~erg~s$^{-1}$ (0.5--10 keV) 
was a factor $\sim$3 smaller than that observed in the 1993 \rosat\ and \asca\  data \citep{Kohno2000}.

The new   period value, P$_{\rm spin}$=2.762383(5)~s,
implies an  average  spin-up rate of $\dot{P}_{\mathrm{spin}} = -(1.27\pm0.01)\times10^{-12}$~s $s^{-1}$ 
in the $\sim$20 years between the two outbursts. 
This  spin-up value is probably not representative of the true torques currently acting on the neutron star, 
since during the  long time interval between the two outbursts,
the pulsar could have been spinning-down, at least for part of the time. 
With the reasonable assumption that an accretion disk is present during the observed outburst, 
we can discuss the results in the framework of the \citet{Ghosh1979} theory of 
accretion torques. 
For a given neutron star mass, radius, and dipolar magnetic field, 
the spin period derivative, $\dot{P}$, is expected to scale as 
$PL$$^{3/7}$ (where $P$ is the spin period, $L$ is the X--ray luminosity).
Assuming $M=1.4\,M_\odot$, $R = 10\,\mathrm{km}$, $L=10^{38}$~erg~s$^{-1}$, 
$B=10^{12}\,\mathrm{G}$ 
the expected spin-up rate for \src\ during the \xmm\ observation is $\dot{P}$$\sim$$-6.8\times10^{-11}$~s $s^{-1}$ 
(eq. 15 in \citealt{Ghosh1979}).
As expected, this  value is larger (by a factor of $\sim$50) than the time averaged spin-up rate between 1993 and 2014.

Although the pulsed fraction observed in 2013   ($8.9\pm0.5$~\%) 
was smaller than in  1993 (37\%), 
the large number of counts collected by the three EPIC cameras allowed us to see interesting variations of the pulse 
profile as a function of energy that were not evident in  the previous observations. 
The pulse profile is  double-peaked  at intermediate energies  (0.5--4 keV) and for the first time
pulsations were seen also  below 0.5~keV. 
In principle, the presence of pulsations in the soft energy range might give some information on 
the origin of the soft excess (see below), but  in practice this is complicated by the fact that  
both the soft excess and the hard power law contribute to the counts in this energy range 
and could be responsible for the observed modulation.

Our observation confirms the presence of a soft excess in \src . 
Its 0.2--12 keV spectrum 
could be well fitted by a double-component continuum with
a hard cut-off power-law ($\Gamma$$\sim$0.44) together with a soft excess equally well modelled by
either a soft black-body (kT$\sim$0.1 keV), or a hot thermal plasma model (kT$\sim$0.2 keV). 
While these spectral parameters are similar to what found in 1993,
the relative contribution of the soft component to the total luminosity
was much lower in 2014: the 0.1--10~keV luminosities of the soft and hard 
spectral components were L$_{\rm bb}$=10$^{36}$~erg~s$^{-1}$
and L$_{\rm cpl}$=7.1$\times$10$^{37}$~erg~s$^{-1}$  in 2014, 
while they were L$_{\rm soft}$=8$\times$10$^{37}$~erg~s$^{-1}$
and  L$_{\rm hard}$=1.8$\times$10$^{38}$~erg~s$^{-1}$ in 1993,  implying a 
soft to hard luminosity ratio of 1.5\% (in 2014) instead of 44\% (in 1993).  

We discovered several features in the \src\ spectrum: 
narrow lines  from ionized oxygen and nitrogen (in the RGS data), 
a broad feature at 0.9-1 keV, likely produced by Fe-L shell emission (both in the RGS and EPIC data), 
and an iron emission line at $\sim$6.6~keV.
The latter is clearly evident in the time-averaged spectrum. 
There is some evidence that its centroid energy varies as a function of the pulsar spin phase, but 
the significance of the line is marginal ($<3 \sigma$) when it attains the highest energy.

We detected a phase dependence of the cut-off energy (from $E_c$$\sim$20~keV to E$_c$$>$45~keV), 
which is the main responsible for the spectral variability with the spin phase.
A similar behavior of high-energy cut-off is unusual in X--ray pulsars (an example is the 
low-mass X--ray binary pulsar 4U 1626-67,  \citealt{Coburn2002}). 
We observed also a phase dependence of the flux contributed by the broad emission component at 0.9-1 keV,
that can be explained by a variable emitting area illuminated/photoionized by the pulsar beam.

An accretion disc is a natural reprocessing site for the X-ray emission emitted from the accreting pulsar, 
especially when its  luminosity is high  ($\sim10^{38}$~erg~s$^{-1}$)
and photons preferentially escape in a fan-beam from the accretion column (e.g. \citealt{Parmar1989}),
favouring an X--ray illumination of the equatorial regions.
Following \citealt{Hickox2004}, we can assume that  black-body emission
with luminosity L$_{SOFT}$=$\Omega$ $L_{\rm X}$ 
(where $L_{\rm X}$  is the total X--ray luminosity)
results from  reprocessing by 
optically thick material located at a distance $R$
and  subtending a solid angle $\Omega$ with respect to the central X-ray source. 
Independent on the exact geometry of the reprocessing region,
the distance $R$ can be estimated from the relation  $$R^2 = \frac{L_{\rm X}}{4 \pi \sigma T_{\rm BB}^4}$$
(note that this is different from the   radius ($r_{bb}$) 
resulting from the normalization of the black-body component in the spectral fit (Tables~\ref{tab:av_spec} and   \ref{tab:bbpow})).
If the reprocessing region is a shell at the inner edge of the disc, 
$R$  should be of the order of the magnetospheric radius, $R_{m}$ (or of the corotation radius  if the pulsar is 
rotating close to the equilibrium period).
Adopting the \src\ phase-averaged luminosity L$_{\rm X}$=7$\times$10$^{37}$~erg~s$^{-1}$ (0.5--10 keV) and 
the black-body temperature, $T_{\rm BB}$, of 0.09~keV  
(Table~\ref{tab:av_spec}), we obtain a distance $R=3\times$10$^{8}$~cm, which is indeed very
similar to the corotation radius $R_{\rm cor}$=3.3$\times$10$^{8}$~cm. 
An estimate of  the  magnetospheric radius  \citep{Davies1981}   can be obtained from the relation
$$R_{m}\sim1.5\times10^{8}m_1^{1/7}R_6^{10/7}L_{37}^{-2/7}B_{12}^{4/7} {\rm cm,}$$
where  $m_1$ is the neutron star (NS) mass in units of solar masses, $R_6$ is the NS radius in units of $10^6$~cm, 
$L_{37}$ is the X--ray luminosity in units of $10^{37}$~erg~s$^{-1}$, $B_{12}$ is the NS magnetic field in units $10^{12}$~Gauss.
If for \src\ we assume $B_{12}$=1 we obtain $R_{m}$$\sim$10$^{8}$~cm.
If the soft excess is caused by reprocessing of the X-ray pulsar emission by the inner edge of the accretion disk, 
some pulsations in the soft part of the spectrum are expected, as indeed it was found.
In this scenario,  the narrow lines we discovered at low energies in the RGS spectrum 
and the ionized iron line at 6.6~keV can be produced by a  plasma located in the disc atmosphere. 
They require the presence of matter with very different ionizations stages.
Indeed, while plasma with highly ionized oxygen and nitrogen is characterized 
by a ionization parameter $\xi$$\sim$100~erg~cm~s$^{-1}$
($\xi$=$L$/$nR$$^{2}$, where $L$ is the luminosity of the ionizing 
source, $n$ is the plasma density, $R$ is the distance of the photoionized
matter from the ionizing source), the presence of highly ionized iron \ion{Fe}{xxv} 
implies  $\xi$=1000~erg~cm~s$^{-1}$ \citep{Kallman1982}.
The widths we derived for the  \ion{N}{vii} and \ion{O}{viii} Ly$\alpha$ lines are larger than the RGS energy resolution 
and imply 
velocities of $\sim$1,000-2,500~km~s$^{-1}$ for the emitting plasma, as expected for 
matter rotating at distances of a few $\times$10$^{9}$~cm from the neutron star.
Given the high X--ray luminosity, the presence of the accretion disc, and the detection of narrow lines from matter with
very different ionization stages not compatible with a single plasma temperature, 
we can interpret  the soft excess as the result of reprocessing  
from the inner edge of the disc and the narrow lines as due to photoionized plasma in regions above the disc.

An  alternative interpretation is that the soft excess in \src\ is entirely emitted by a thermal hot plasma \citep{Hickox2004}.
Using a {\sc mekal} model to  fit the RGS spectrum, we could account well   {\em only} for the 
\ion{O}{viii} Ly$\alpha$ emission line, which
we interpret as  a further indication of the presence of  photoionization, and
obtained a metal abundance 
($Z \lesssim 0.01$ $Z_{\sun}$)  too low for the SMC. 
The emission measure
obtained from the normalization of the {\sc mekal} model 
is  $n^2V \sim 10^{60}$~cm$^{-3}$. Since the gas density in this case should be   $n < 10^{12}$~cm$^{-3}$,
the emitting spherical region for the optically thin plasma should have a radius  $R_{\rm cloud} > 6\times 10^{11}$ cm,
and no pulsations in the soft excess are expected.

\citet{Kohno2000} found wavy structures below 1~keV in their \rosat\ plus \asca\ spectrum 
and explained them with the presence of a deep edge-like structure at 0.5~keV, caused
by an over-abundance of  neutral oxygen (about 6 times solar) in the circumstellar matter.
However, this was at odds with the very low metal abundance (0.02 solar)
in their best-fit thermal plasma model.
These authors concluded that the oxygen over-abundance is real, while the {\sc mekal} model is un-physical.
Also our \xmm\ spectrum showed wavy structures below 1 keV, 
which we fitted with a broad Gaussian line at $\sim$0.96~keV, but we  
found no evidence for a  neutral oxygen edge at 0.54~keV deeper than the one ascribable to the best-fitting N$_{\rm H}$.
We note that acceptable fits to the EPIC spin-phase selected spectra adopting 
a single absorption model with variable
metal abundances  ({\sc vphabs} in {\sc xspec}) could be obtained {\em only} deleting the Gaussian component at 0.96~keV.
However, in this case, an oxygen overabundance around 30 times solar was obtained, which we think is unlikely.
Moreover, the resulting column density was very low  ($\sim$5$\times$$10^{19}$~cm$^{-2}$),
one order of magnitude lower than the Galactic one.
If, instead, we use two absorption models, one for the Galactic   absorption ({\sc phabs}) 
and  one  with variable abundances for the SMC absorption ({\sc vphabs}),
the   column density of the latter becomes compatible with zero and the fit is insensitive to the metal abundances.
Given that the neutral oxygen edge falls within two bright emission lines in the RGS spectrum 
(\ion{N}{vii} and \ion{O}{viii}), we believe that  the edge-like structure at 0.5~keV observed by \citet{Kohno2000} 
was due to the low spectral resolution of the \rosat\ plus \asca\ spectra  and to 
the presence of the nearby  emission component at 0.9-1~keV.

A hint of an absorption line at 7.5~keV was present
only in a narrow range of spin phases  ($\Delta$$\phi$=0.7-0.8). 
Since its energy is not consistent with any known line feature (except for the K$_{\alpha}$ line from neutral Ni at 7.48~keV,
which we consider very unlikely, given the highly ionizing flux from the X--ray pulsar), 
it is tempting to associate it with a cyclotron resonant scattering feature. 
In this case, its energy would imply a  magnetic field in the scattering region
of 6.5$\times$(1+z)$\times$10$^{11}$~G, where z is the gravitational redshift. 
%

\section{Conclusions}

The main results of 
our  \xmm\ observation obtained during  the second bright outburst from the SMC Be/XRT \src\ 
can be summarized as follows :

\begin{itemize}

\item  The continuum  X--ray spectrum is complex and requires several components: a hard power law with a
phase-dependent high-energy cut-off,  a soft excess below 0.5 keV well fit by a black-body model, 
and a broad emission feature at $\sim$0.9-1~keV.

\item We discovered narrow emission and   absorption lines 
from  oxygen, nitrogen and iron in different ionization stages.

\item The relative contribution of the soft excess component to the total luminosity was smaller than in the 
1993 observations (1.5\% wrt 44\%), when the source was a factor $\sim$3 brighter.

\item Contrary to what observed in 1993, the pulsar pulse profile was strongly energy-dependent and
pulsations were detected also below 0.5~keV. The pulsar spin period of P$_{\rm spin}$=2.762383(5)~s implies an average spin-up rate
of $\dot{P}_{\mathrm{spin}} = -(1.27\pm0.01)\times10^{-12}$~s $s^{-1}$ in the 
$\sim$20 years elapsed after the only other period measurement of this source.

\end{itemize}

These findings lead us to interpret the soft X-ray excess in this source as due to reprocessing of the
pulsar emission from  (part of) the inner edge  of the  accretion disc at a radius of $\sim3\times10^8$ cm. 
The disc inner boundary is consistent with a magnetic field of the order of 10$^{12}$ G which could also explain the
possible absorption feature at 7.5~keV  as an electron cyclotron line. 
The observed narrow lines are most likely produced  by photoionization of plasma located above the inner regions of the accretion disc.

\section*{Acknowledgments}

This work is based on data from observations with \xmm.
\xmm\ is an ESA science mission with instruments and
contributions directly funded by ESA Member States and the USA (NASA).
LS thanks A.~Pollock and  A. Gim{\'e}nez-Garc{\'{\i}}a for interesting discussions.
PE acknowledges a Fulbright Research Scholar grant administered by the U.S.--Italy Fulbright Commission 
and is grateful to the Harvard--Smithsonian Center for Astrophysics for hosting him during his Fulbright exchange.
We acknowledge financial contribution from the agreement ASI-INAF I/037/12/0.

\begin{appendix}

\section{MOS2 timing mode}
\label{app:m2}

Here we discuss some calibration problems that affected the MOS2 spectrum, which was not considered
in the spectroscopy of \src. Note however that these issues
have no impact on the timing analysis, which was indeed performed using also MOS2 events.

Source counts in MOS2 timing mode were initially extracted from RAWX=290 to RAWX=321, 
resulting in a count rate $\sim$5~count~s$^{-1}$. 
Background was taken from an outer CCD that collected data in imaging mode.
When compared with the spectra extracted from the other cameras, MOS2 spectrum
showed a significant deficit of counts above 7 keV, a shortage becoming much less evident when 
extracting counts from a box excising the central columns of the PSF (that peaks at RAWX=305).
We extracted different spectra from boxes with a variable number of excluded inner columns.
For each  box region, we also tried different PATTERN (single-pixel events  or single and double events), 
but this led to a marginal difference in the resulting spectral shape in the energy range 0.6--10 keV.
We followed the procedure recommended on the \xmm\ {\sc sas} User Guide (issue 10.5, updated to 2014, February 7) available at
{\sc http://xmm2.esac.esa.int/docs/documents/} to properly 
calculate the effective area associated with a spectrum extracted from 
the box in MOS2, from which the core columns have been excised.
Comparing these several different MOS2 spectra, and fitting them with a simple power law model, we
found significantly different slopes, even when restricting the spectral analysis to the safer energy range 2-7 keV.
The reason for this behaviour is unclear (we did not find anything similar reported in the {\sc sas} User Guide).
In principle, the count rate is too low ($\sim$5~count~s$^{-1}$) 
to cause possible pile-up problems in MOS2, 
although the use of the {\sc sas} task {\sc epatplot} on MOS2 events (for both the whole spatial RAWX region and the different
boxes chosen) showed a significant discrepancy 
in the distributions of observed and expected event shapes
in the energy range 0.5-2 keV, while apparently no problems were found above 2 keV. 
We also noted a significant electronic noise in MOS2 image (TIME versus RAWX), with a reapeting pattern
in RAWX (several equally spaced ``hot RAWX columns''), appearing only below 0.5~keV.
It is possible that this huge noise creates significant problems in the MOS2 spectral reconstruction.

\end{appendix}

\bibliographystyle{mn2e} 
\bibliographystyle{mnras}

\bsp

\label{lastpage}

\end{document}